\documentclass[journal=langd5,manuscript=article]{achemso}

\usepackage{chemformula} % Formula subscripts using \ch{}
\usepackage[utf8]{inputenc}
\usepackage[T1]{fontenc} % Use modern font encodings
\usepackage{chemist}
\usepackage{color}
\graphicspath{{figures/}}

\title{FM-AFM with hanging fiber probe for the study of liquid / liquid interfaces}

\author{Matthieu Rocheron}
\affiliation[Aix Marseille Université]
{Aix Marseille Université, CINAM UMR 7325, F-13009 Marseille}
\author{Christian Curtil}
\affiliation[Aix Marseille Université]
{Aix Marseille Université, CINAM UMR 7325, F-13009 Marseille}
\author{Hubert R. Klein}
\email{hubert.klein@univ-amu.fr}
\affiliation[Aix Marseille Université]
{Aix Marseille Université, CINAM UMR 7325, F-13009 Marseille}

\abbreviations{QTF, FM-AFM}
\keywords{liquid liquid interface, FM-AFM, rheology, surface tension}

\begin{document}

\begin{tocentry}
\begin{center}
 \includegraphics[scale=0.1]{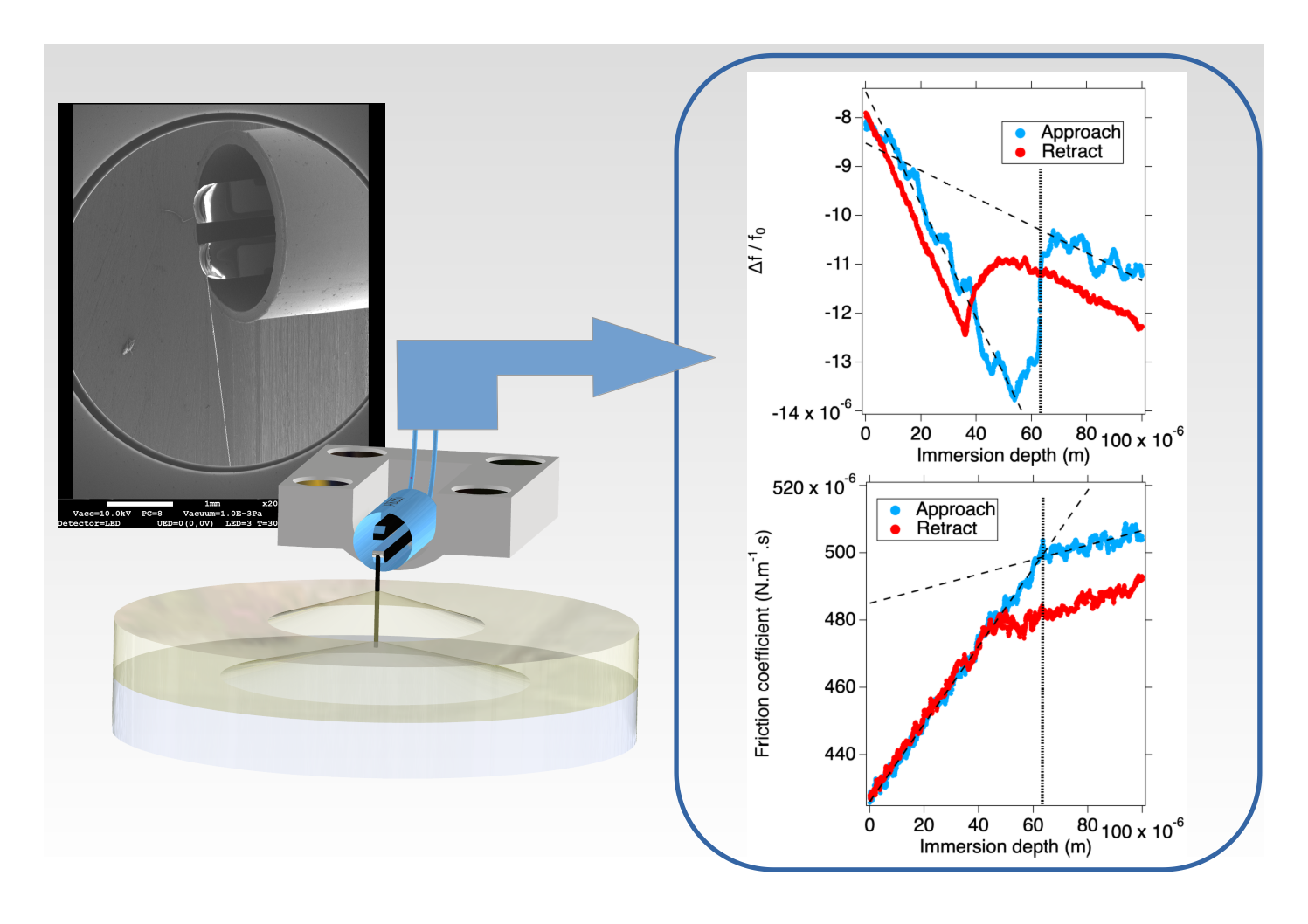}
\end{center}
\end{tocentry}

\begin{abstract}
This article describes how a frequency modulation AFM using a hanging fiber force probe made from a quartz tuning fork provides local measurements on liquid-liquid interfaces. After detailing the manufacture and calibration of the force probe, we provide evidence that this AFM is suitable for quantitative measurements at the interface between two liquids. The repeatability of the measurements allows a poly-dimethylsiloxane / water moving interface to be monitored over several hours : the evaporation of a water droplet immersed in poly-dimethylsiloxane is observed, and its interfacial tension evolution over time is measured. Deformation of the interface is also observed. These capabilities, and preliminary results on the interface between two immiscible liquids, pave the way for interface manipulation and study of complex fluid-fluid interfaces.
\end{abstract}

\section{Introduction}

Understanding and engineering of liquid interfaces has interested both fundamental science and industry for decades \cite{degennes2004}.Interfaces between two immiscible liquids are exploited to control nucleation \cite{Grossier2011} and crystallization \cite{Bonnett2003,Galkin2000}, for droplet generation in microfluidic devices \cite{Doufene2019}, and for asymmetric chemical synthesis \cite{Hassan2015}. They also serve as templates for the self-assembly of nanoparticles \cite{Turek2012}. Many of these applications manipulate small volumes, which means that liquid interfaces need to be probed at a local scale. Atomic Force Microscopes (AFM) are undoubtedly good candidates for the job.

Since their invention, AFMs have been widely used in liquid media. For example, they have opened the way to the exploration of wetting and capillarity on a local scale \cite{Xiao2000,DupredeBaubigny2015}, of surface forces between a colloid and a liquid interface\cite{Hartley1999}, of the structure of liquids close to a surface \cite{Zhmud1998}, or even the imaging of living cells \cite{Haberle1991}. During the last decade, they have been applied to the imaging of bubbles, drops \cite{Uddin2011,Munz2014}, nanoparticles or self-assemblies at the interface of two liquids \cite{Costa2016}, but also to the study of the wetting dynamics of coated air / water interfaces \cite{Guo2021}. At the same time, several teams have successfully studied the possibility of using an AFM as a microrheometer \cite{Mechler2004,Devailly2014,AguilarSandoval2015}. 

Working in a liquid environment greatly reduces the sensitivity of standard oscillating AFM probes. The probes have to be completely immersed, which results in over-damping of the oscillator\cite{Munz2014,Uddin2011}. Moreover, their tapered shapes hamper the interpretation of rheological measurements. This led to the development of near constant diameter dedicated probes for AFM rheological measurements, in the form of a whisker \cite{Mechler2004}, or hanging cylindrical fiber \cite{Xiong2009}. These oscillating probes require only the whisker or the fiber to be immersed in the liquid, which solves the problem of over-damping. Sensitivity can be further improved by using quartz tuning forks (QTF) as oscillators instead of standard AFM cantilevers.

The use of quartz tuning forks as force probes was envisaged very early on by the pioneers of AFM\cite{Gunther1989}. QTFs are high quality factor ($Q \simeq 10^4$) and stiff ($k\simeq 10^4$N.m$^{-1}$) oscillators. The quality factors of these oscillators are much higher than for standard AFM probes, which gives them unrivaled force sensitivity as low as a few tenths of pN.Hz$^{-1/2}$. Coupled with their high static stiffness in the range of $10^4$N.m$^{-1}$ \bibnote{at the cost of a limited measurement bandwidth}, this makes them very suitable for work in liquids and/or at soft interfaces\cite{Tamayo2001}. Moreover, operated in Frequency Modulation mode (FM-AFM ), they allow independent measurement of the conservative and dissipative parts of the interaction\cite{Albrecht1991}.

We report here how a quartz tuning fork (QTF) in a hanging fiber geometry (i.e. a cylindrical fiber glued to a prong of the tuning fork, and oscillating along its major axis) was used to perform AFM measurements at the interface between two liquids. We describe the dedicated AFM and probes developed for this purpose, and the principle of FM-AFM measurements in liquids. Validation of the technical feasability of this setup is obtained through quantitative rheological measurements in simple fluids. Finally, we present the first results of measurements at the interface between two immiscible liquids, poly-dimethylsiloxane (PDMS) and water.

\section{Experimental}

\subsection{Experimental setup}

\begin{figure}
 \includegraphics[scale=0.25]{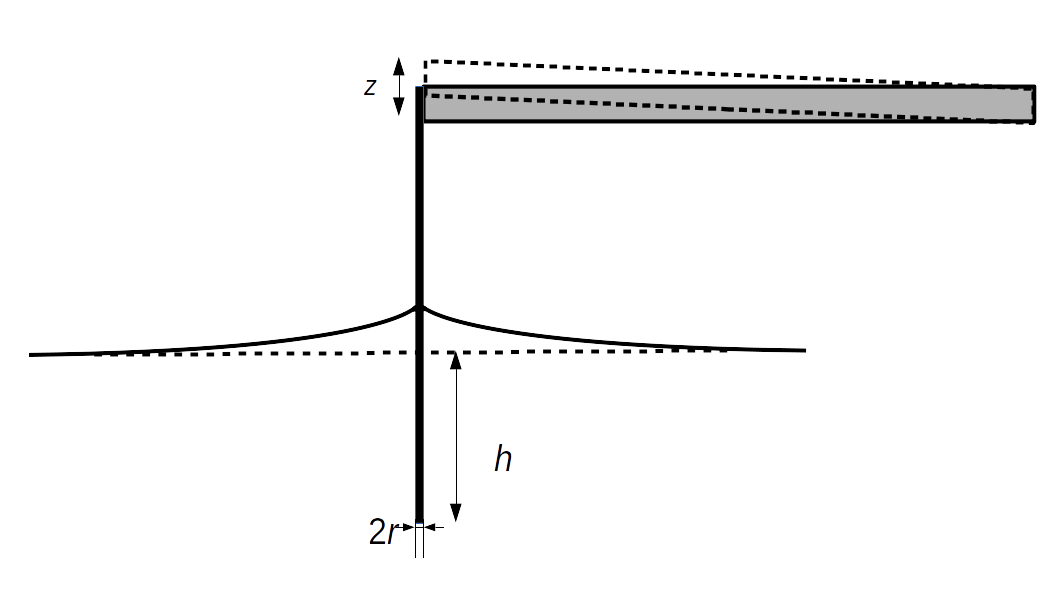}
 \includegraphics[scale=0.5]{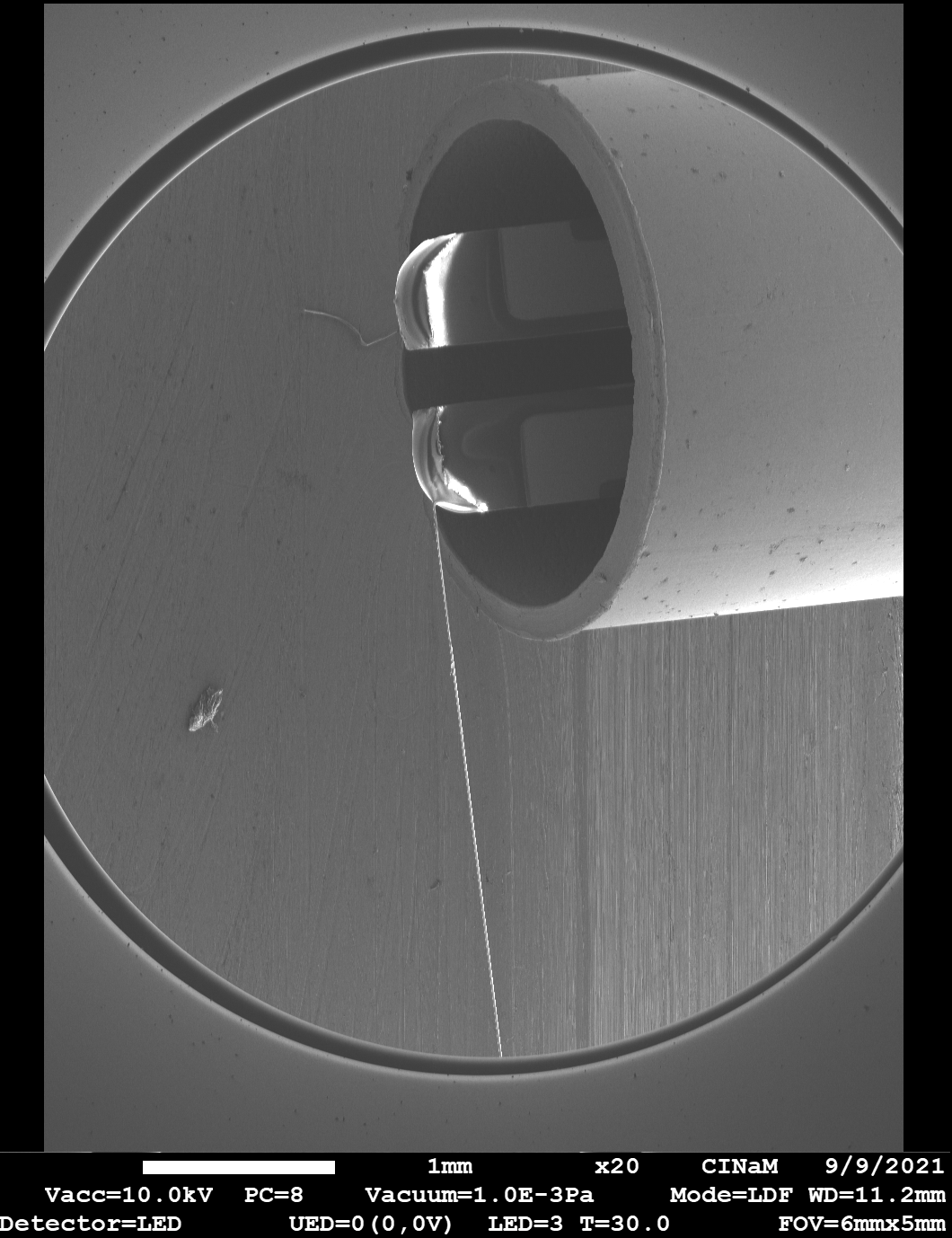}
  \caption{\textbf{Left :} Schematic diagram of the experiment. One end of a fiber of radius $r$ glued to a prong of a quartz tuning fork is dipped from a height $h$ above the the air - liquid interface. It oscillates along its axis with an amplitude $z$. The frequency and oscillation amplitude of the QTF prong are monitored. \textbf{Right :} SEM image of a fibered QTF. The prongs of the QTF protrude from the cut packaging. A glass fiber thinned to a radius $r$ is glued perpendicularly to the end of one prong, and the same amount of glue is applied to the other prong to balance the mass. The fiber radius is $r=4.5\mu$m.}
  \label{sch:qtf_head}
\end{figure}

Measurements were performed with a glass fiber glued to a quartz tuning fork in a hanging fiber geometry, where the fiber oscillates along its long axis and is partially dipped in the fluid (see Figure \ref{sch:qtf_head}). The advantage of this method is that the oscillator is not damped by the liquid, allowing precise measurements of the fiber's interaction with the liquid \cite{Devailly2014,Xiong2009,Jai2006}. 

The QTF probe was used in a home made AFM mounted on an inverted optical microscope (TE2000-U, Nikon).  The fluid is contained in a liquid cell with a diameter of 30 mm and a depth of 3 mm, the bottom consisting of a glass cover slip changed for each new experiment. The fluid sample  can be scanned in the three directions using a piezoelectric actuator (P527-3CL, Physik Instrumente) positioned under the liquid cell and operated in closed-loop mode . Scanning is performed by an open source digital signal processing controller (MK2-A810, Soft dB), driven by the GXSM software \cite{Zahl2010}.  The QTF holder (containing a piezo dither), positioned above the liquid cell  is mounted on a vertical 22mm piezoelectric slider operated in closed-loop mode (Q521, Physik Instrumente). It offers a bidirectional repeatability of 80nm for controlled long range vertical movements in fluids. The QTF is excited mechanically by a piezo dither, in mechanical contact with its base, and its oscillation amplitude is measured using a differential voltage amplifier (DLPVA-100, 40dB gain, 100kHz bandwidth, Femto) to ensure an optimal signal to noise ratio \cite{Jahncke2004}. Furthermore, the 1T$\Omega$ input impedance of this amplifier eliminates the need to compensate for the stray capacitance of the tuning fork \cite{Grober2000}, as virtually no current flows through it.

The AFM is operated in the FM-AFM mode using a phase-locked-loop device (PLL, HF2LI, Zurich Instruments) driven by the LabOne software. In this mode, the QTF oscillates at its resonance frequency, and the frequency shift $\Delta f$ with respect to the natural resonance frequency in air is monitored. A proportional-integral-differential (PID) closed loop is used to modulate the excitation voltage $A_{ex}$ sent to the piezoelectric dither, in order to maintain the amplitude of oscillation constant. Monitoring the $A_{ex}$ signal gives access to the dissipation of the oscillator (see FM-AFM measurements section for details).

The whole setup is operated on an optical table in a basement, to ensure minimum mechanical vibrations and temperature variation. It enables the probe to be positioned above a liquid interface, and allows control of the dipping depth relative to the air-liquid interface level. The setup also allows  independent measurement of the conservative and dissipative parts of the interaction between the fiber and the fluid.

\subsection{Building and calibration of the FM-AFM probe}

Our hanging fiber probe was produced from a quartz tuning fork. The quality factor is defined as $Q=f_0 / \Delta f$ where $f_0$ is the resonance frequency of the QTF and $\Delta f$ its full frequency width at half maximum (which we will note FWHM in the following) of the tuning fork amplitude $<z>$ at resonance.  

The aluminium can containing the QTF, sealed under residual vacuum, was first cut sufficiently  to make the prongs of the tuning fork accessible, using a watchmaker's lathe. The fiber, a pure silica optical fiber (initial diameter 125$\mu$m) was chemically thinned down in 30\% hydrofluoric acid to the desired radius \cite{Kbashi2012}, typically $r=5 \mu$m for this study. The thinned fiber was glued to the top of one prong of the QTF and perpendicular to it (see the micrography of figure \ref{sch:qtf_head}) using a UV curing glue (Nordland optical adhesive 81). It should be noted that the glue is applied at a temperature of 5°C, this gives it the texture of a gel during application and prevents the glue from wetting the entire fiber and thus modify its surface.

 The fiber was then cut to the desired length (typically 5mm). To avoid a strong decrease in the quality factor of the oscillator, the masses of the two prongs need to be be as similar as possible;  a glue droplet was therefore deposited at the top of the other prong of the QTF. All these operations were performed using micromanipulators under a binocular microscope. This protocol typically led to QTF probes with a resonance frequency in air of around 32 kHz, and a quality factor in the range of  $10^4$.

Using a QTF probe as a force probe require determining not only its spring constant, but also its electrical sensitivity, so as to control the oscillation amplitude. This was because we were measuring the piezo-electrical response of the quartz to the oscillation amplitude of the QTF prongs. The QTF was modeled here as two coupled oscillators to avoid underestimation of the spring constant. The effective spring constant $k_{eff}$, and effective mass $m_{eff}$ were estimated by a geometrical analysis of the prong's dimensions and by measuring of the frequency of the in-phase ($f_i$) and anti-phase ($f_a=f_0$, the resonance frequency) oscillation modes of the QTF, as described by Castellanos-Gomez \textit{et al} \cite{Castellanos-Gomez2009}. The electrical sensitivity $\alpha$ was then determined by measuring the thermal noise spectrum of the bare QTFs. These procedures are described in the Supplementary information, and the results are summarized in Table \ref{tbl:qtf_calibr} for the two kinds of QTFs used in this study. We used QTFs in the form of radial cylinders measuring 8 x 3mm (QTFA), and 6 x 3mm (QTFB) respectively.

\begin{table}
  \caption{Summary of stiffness and electrical sensitivity calibration of bare QTFs. Calibration measurements were performed at T=22°C.}
  \label{tbl:qtf_calibr}
  \begin{tabular}{lll}
  	  & QTFA & QTFB \\
    \hline
    $L$ ($\mu$m)   & $4000 \pm 120 $ & $3200 \pm 96$   \\
    $T$ ($\mu$m) & $590 \pm 12$ & $400 \pm 7 $\\
    $W$  ($\mu$m)& $350 \pm 7$ & $330 \pm  7$ \\
    $f_i$ (Hz) & 16000& 17300 \\
    $f_0$ (Hz) & 32764 & 32757 \\
    $Q$ & 13930 & 9200\\
    $k_{eff}$ (N.m$^{-1}$) & $8.8 (\pm 0.3).10^4$ & $4.7 (\pm 0.16).10^4$ \\
    $m_{eff}$ (kg) & $2.26 (\pm 0.11).10^{-6}$ & $1.12( \pm 0.06).10^{-6} $\\
    $\alpha$ ($m.V^{-1}$) & $4.9(\pm 0.7).10^{-8}$ &  $6.9(\pm 1.1).10^{-8}$ \\
    \hline
  \end{tabular}
\end{table}

\subsection{Materials}

Liquid poly-dimethylsiloxane (PDMS) trimethylsiloxy terminated, \chemform{(CH_3)_3SiO(Si(CH_3)_2O)_nSi(CH_3)_3} was purchased from  Alfa  Aesar. It  was used as obtained: molar mass 1250 g.mol$^{-1}$, density 0.935 and kinematic viscosity 9-11cSt at 25$^\circ$C. 

The water used for this study was deionized with a resistivity of 18M$\Omega.$cm.

The QTFs were watch quartz crystals, of nominal resonant frequency 32768 Hz. We used radial cylinders measuring 8 x 3mm (QTFA) and 6 x 3mm (QTFB) with a 12.5 pF load capacitance , purchased from AEL CRYSTALS ( references X32K768L104 and X32K768L009 respectively ).

The QTF probes were built using single mode fibers with a cladding diameter of $125\mu$m and a pure silica core of diameter $3.5\mu$m, dual acrylate coated, purchased from Thorlabs. The coating was removed by dissolution in dichloromethane.

\subsection{FM-AFM measurement principle}

\begin{figure}
  \includegraphics[scale=0.25]{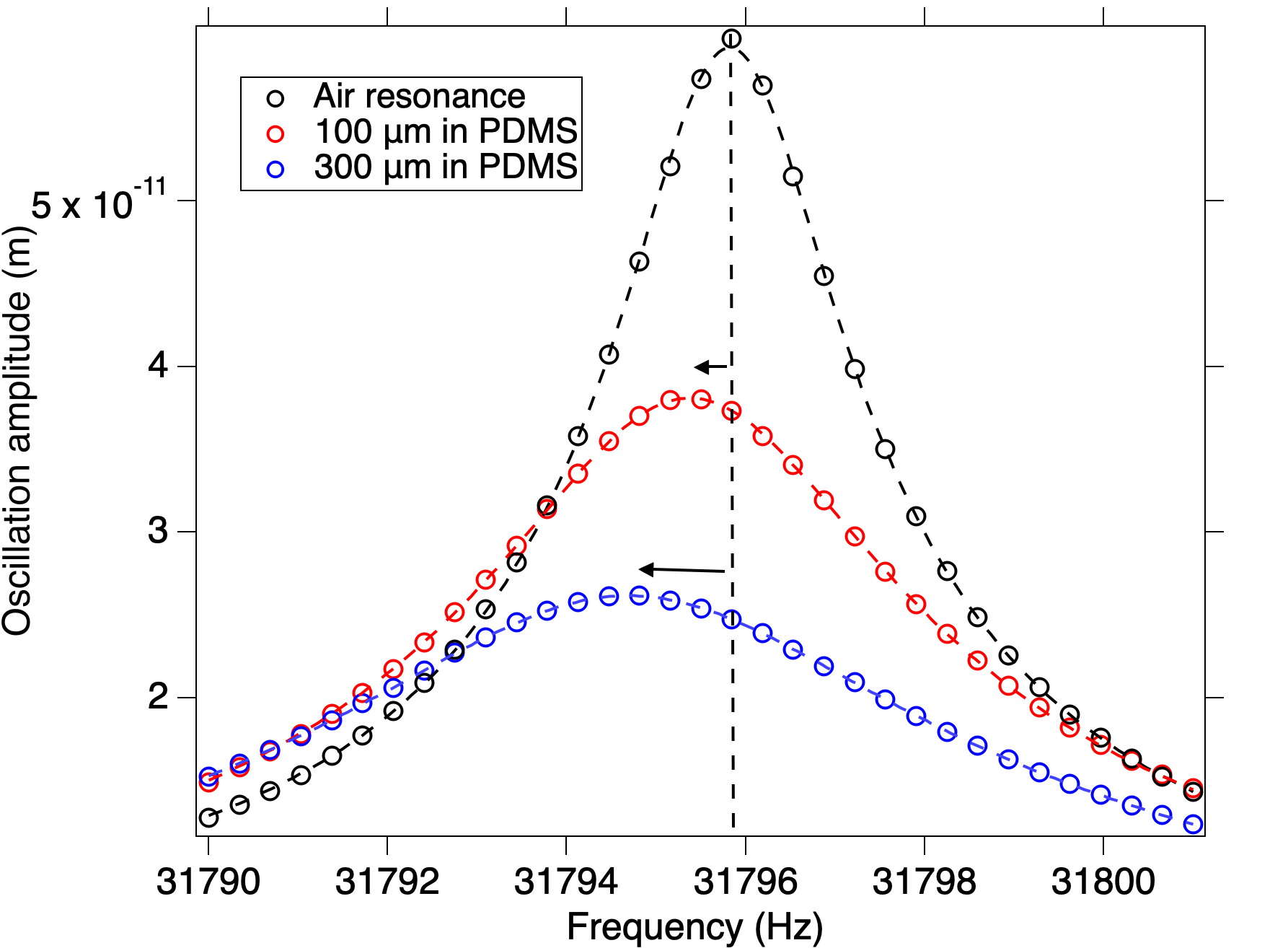}
  \caption{Resonance spectra of a fibered QTFA (fiber radius $r=4.5 \mu$m) in air and with the fiber immersed at different depths in 10cst PDMS oil. The dashed lines are the best fits using a Lorentzian function. Modeling yields a resonant frequency $f_0=31795.8$Hz, and a quality factor $Q=12950$ in air. As the length of the immersed fiber increases, resonance frequency decreases due to increased drag mass, and FWHM increases due to increasing dissipation within the fluid.}
  \label{sch:qtf_resonance_pdms}
\end{figure}

Quantitative rheological measurements necessarily involve measuring the conservative and dissipative parts of the fiber's interaction with the fluid. These measurements are obtained by comparing the resonance spectrum of the oscillator, compared to the spectrum in air \cite{Devailly2014}. Figure \ref{sch:qtf_resonance_pdms} shows three resonance spectra recorded in air and at two different depths in a 10cst PDMS bath. Compared to the spectrum recorded in air, the resonance frequency undergoes a negative shift when dipped in liquid, and this shift increases with the dipping depth. 

The fiber, while oscillating in the liquid, puts a layer of fluid in motion at its periphery. The thickness of this moving fluid layer depends on the viscosity and density of the fluid\cite{Tirado1979}, through the skin thickness of a viscous flow induced by an oscillating plane in a liquid\cite{Landau1987}:
\begin{equation}
 \label{eq:delta}
  \delta = \sqrt{\frac{2\eta}{\rho \omega}}
\end{equation}
with $\eta$ the dynamic viscosity (Pa.s) and $\rho$ the density of the fluid (kg.m$^{-3}$), $\omega$ the angular frequency (rad.s$^{-1}$).

Since the resonance of the oscillator is given by 
\begin{equation}
 f_0= \frac{1}{2\pi}\sqrt{ \frac{k_{eff}}{m_{eff}} }
\end{equation}
with $k_{eff}$ and $m_{eff}$ the effective stiffness and mass, the drag of this layer of fluid increases the effective mass of the oscillator, and therefore causing a decrease in the resonance frequency. This decrease in resonance frequency is expected to vary linearly with the length of the fiber immersed.

Meanwhile, the FWHM of the spectra increases, indicating increased dissipation, also related to the density and viscosity of the fluid.

Both the dragged mass and dissipation are expected to vary linearly with the length of the fiber immersed in the fluid. 
Because of the quality factor of the oscillator, acquiring such spectra at appropriate resolution takes a few minutes using a lock-in amplifier. We use an alternative method to obtain the same information faster: servo control of the frequency and amplitude of the oscillations. We use a PLL to maintain the oscillator at its resonance frequency, and monitor the frequency shift $\Delta f$ with respect to the natural resonance frequency in air $f_0$. Under these conditions, the elastic and inelastic responses of the oscillator are in quadrature \cite{Giessibl2003}, and the normalized frequency shift reads

\begin{equation}
\label{eq:freq_shift}
 \frac{\Delta f}{f_0} \simeq \frac{1}{2}\frac{\Delta k}{k} - \frac{1}{2}\frac{\Delta m}{m}
\end{equation}

and gives access to variations in the oscillator's effective stiffness and mass. When the fiber penetrates a fluid, the formation of a meniscus is accompanied by an increase in the stiffness of the oscillator. Then, when the fiber is moved in the fluid, since the stiffness of the oscillator is expected to be constant, an added mass per unit length can be calculated using
\begin{equation}
 \Delta m^* = -2m\left( \frac{\Delta f}{f_0}\right)^*
\end{equation}
where $\left( \frac{\Delta f}{f_0}\right)^*$ is the relative frequency shift per unit length. At an interface between two fluids, wetting phenomena \cite{degennes2004} can result in an additional stiffness \cite{Jai2006} that will also modify the frequency shift. A fuller discussion of this point is contained in the section on investigating a PDMS / water interface.

A PID closed loop maintains the oscillation amplitude at a given setpoint by varying the amplitude of the mechanical excitation $A_{ex}$ with respect to the amplitude of excitation in air $A_0$. From this observable, we can extract a friction coefficient $\beta$ in the form of \citep{Giessibl2003} :
\begin{equation}
 \Delta \beta = \beta _0 \left( \frac{A_{ex}}{A_0}-1\right)
\end{equation}

with $\beta _0 = k/\omega _0 Q_0$ (N.m.s$^{-1}$), the intrinsic friction coefficient of the tuning fork in air, $Q_0$ being the quality factor of the oscillator in air, and $\Delta \beta=\beta-\beta _0$. A friction coefficient per unit length $\beta ^*$ can be measured as the slope of the friction coefficient $\beta$'s evolution with depth.

\begin{figure}
  \includegraphics[scale=0.25]{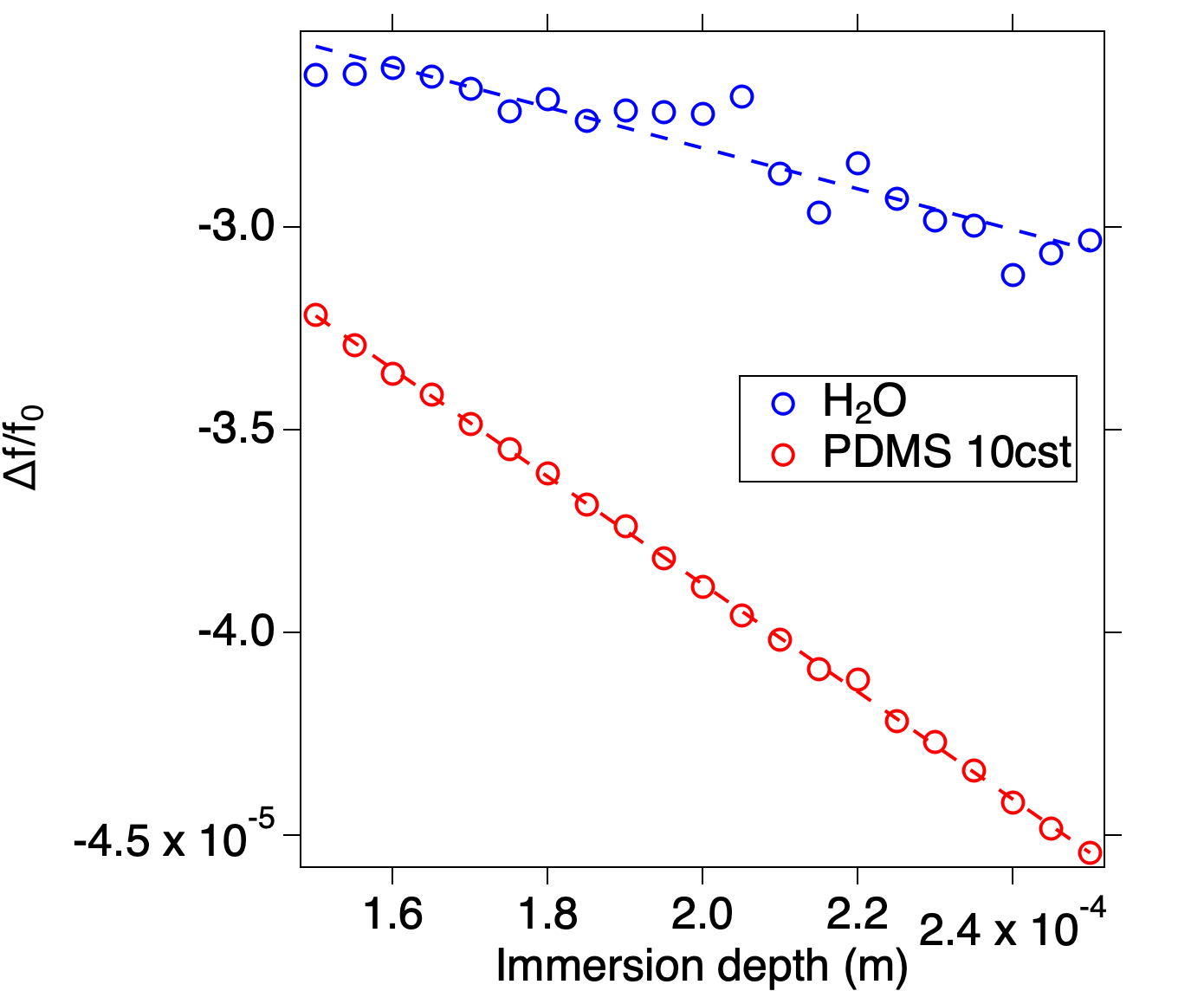}
  \includegraphics[scale=0.25]{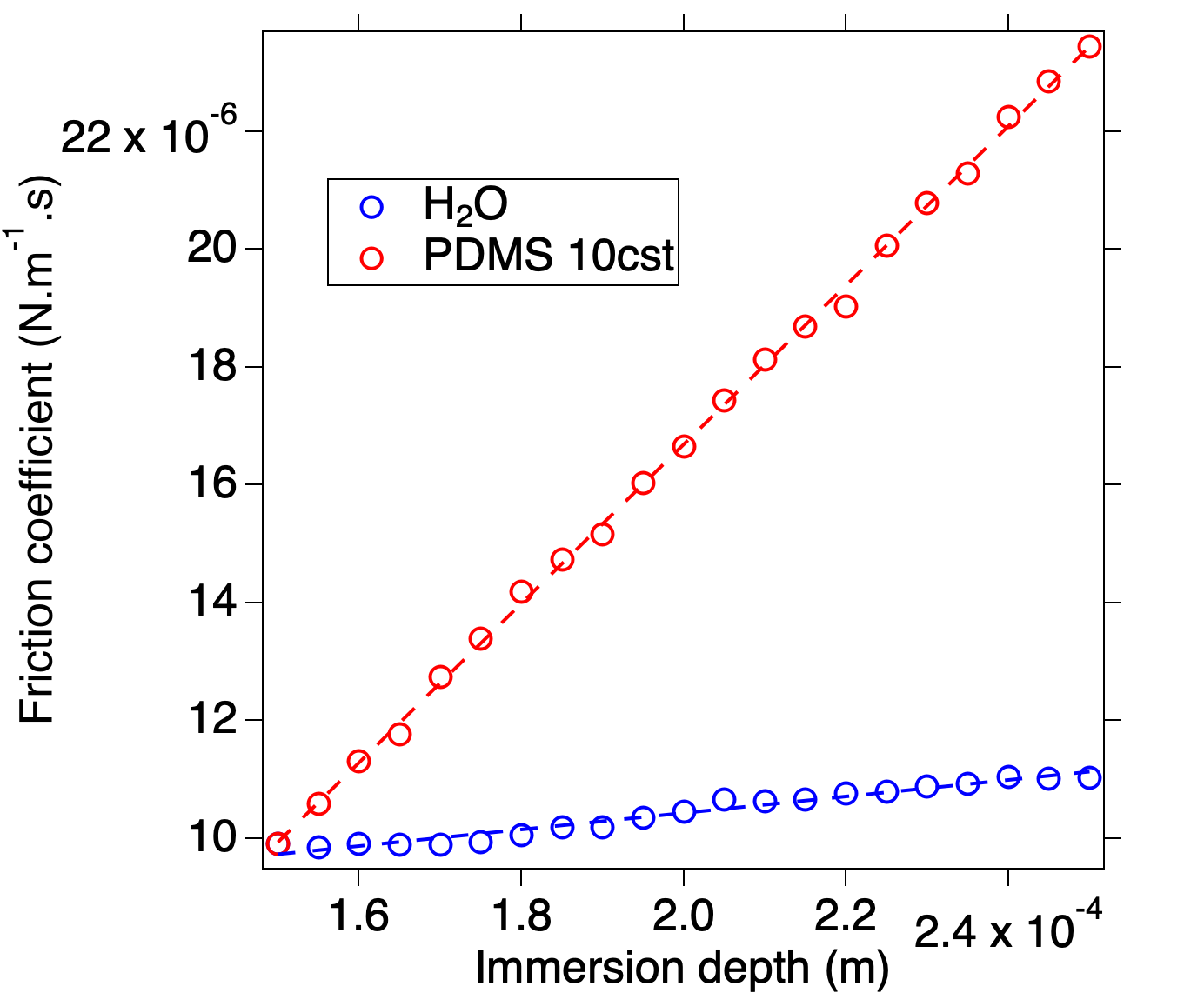}
  \caption{Evolution of the frequency shift (left) and dissipation (right) of a QTFA with respect to the immersion depth of a fiber of radius $r=4.5\mu$m in water or 10 cSt PDMS oil. Frequency shift and friction coefficient vary linearly with the immersion depth, and slopes $\Delta m^*$ and $\beta ^*$ significantly depend on the viscosity of the fluids.}
  \label{sch:df_diss_h2o_10cst}
\end{figure}

Figure \ref{sch:df_diss_h2o_10cst}, shows the evolution of the frequency shift and friction coefficient with respect to the immersion depth of a glass fiber of radius $r=4.5\mu$m in water and 10 cSt PDMS, under downward fiber motion. The recording time for a whole curve is about 5 minutes. As these two values depend on the contact surface between the fiber and the fluids, they vary linearly with the length of the fiber immersed. Their slopes $\Delta m^*$ and $\beta ^*$ obviously vary significantly with the viscosity of the fluids.

\section{Results and discussion}

\subsection{Dragged mass and friction coefficient of simple fluids}

Although the focus of this study was liquid-liquid interfaces rather than high resolution viscosity or density measurements, an important first step was quantitatively measuring the $\Delta m^*$ and $\beta ^*$ coefficients. We thus performed measurements in liquids of various densities and viscosities whose properties are reported in Table \ref{tbl:fluid_properties}. 

\begin{table}
  \caption{Summary of properties of fluids used for mass and friction coefficient measurements at 22° C. Fluids including PDMS oils of various viscosities, a fluorinated oil (GPL103), water and glycerol / water mixtures. Glycerol / water mix viscosities and densities were calculated from a formula proposed by Cheng \cite{Cheng2008}}
  \label{tbl:fluid_properties}
  \begin{tabular}{cccc}
  liquid&viscosity (Pa.s)& density (kg.m$^{-3}$)&$\delta$ ($\mu m$)\\
    \hline
    PDMS 1cst & 8.2.10$^{-4}$ & 820 & 3.15 \\
    Water & 1.10$^{-3}$ & 1000 & 3.15 \\    
    Glycerol / water (0.5/0.5) & 8.4.10$^{-3}$ & 1142 & 8.54 \\
    PDMS 10 cst & 9.35.10$^{-3}$ & 935 & 9.96 \\
    PDMS 100cst & 9.52.10$^{-2}$ & 952 & 31.48 \\
    GPL103 oil & 1.41.10$^{-1}$ &  1900 & 28.53 \\
    Glycerol / water (0.96/0.04) & 6.12.10$^{-1}$ & 1252 & 69.65 \\
    PDMS 1000cst & 9.71.10$^{-1}$ & 971 & 99.56\\
    
    \hline
  \end{tabular}
\end{table}

$\Delta m^*$ and $\beta ^*$ were obtained following the protocol described in the Experimental section, using two different probes 
: a fibered QTFA ($f_0=32116$ Hz, $Q=12950$, fiber radius $r=4.5\mu m$) and a fibered QTFB ($f_0=32110$ Hz, $Q=9600$, fiber 
radius $r=5.5 \mu m$).

Immersion depth was varied over 250 $\mu$m and the vertical speed during measurements was $v=1 \mu$m.s$^{-1}$. The oscillation amplitude of the fiber was set at $z=1$ nm. All measurements were performed at 22°C.

To compare the coefficients measured, which depend both on fiber radius and on the viscosity and density of the fluids, we express them as dimensionless parameters, using the model developed by T. Ondarçuhu \textit{et al} \cite{DupredeBaubigny2016}. This model revisits the one established by G.K. Batchelor for a cylinder moving steadily in a viscous liquid \cite{Batchelor1954}.

Liquid behavior is described by the dimensionless $r/\delta$ parameter, $r$ being the radius of the moving cylinder, and $\delta$ the skin thickness of the viscous flow defined in the previous section.
Dimensionless drag mass $C_m$ and friction coefficients $C_{\beta}$ are expressed as
\begin{eqnarray}
C_m = \frac{r}{\delta}(g_1(r/\delta)-g_2(r/ \delta)) \\
C_\beta\ =  \frac{r}{\delta}(g_1(r/\delta)+g_2(r/ \delta) )
\end{eqnarray}
with the function $g$
\begin{equation}
 g(u) = \frac{K_1[(1-i)u]}{K_0[(1-i)u]}=g_1(u)+ig_2(u)
\end{equation}

where $K_n$ denotes the modified Bessel function of the second kind of order $n$,  $g_1$ and $g_2$ being the real and imaginary parts of $g$.

The drag mass $\Delta m^*$ and friction coefficient $\beta ^*$ per unit length introduced in the previous section are then expressed as :
\begin{eqnarray}
\Delta m^*  = 2\pi \frac{\eta}{\omega} C_m \\
\beta ^* =  2 \pi \eta C_\beta
\end{eqnarray}

\begin{figure}
  \includegraphics[scale=0.25]{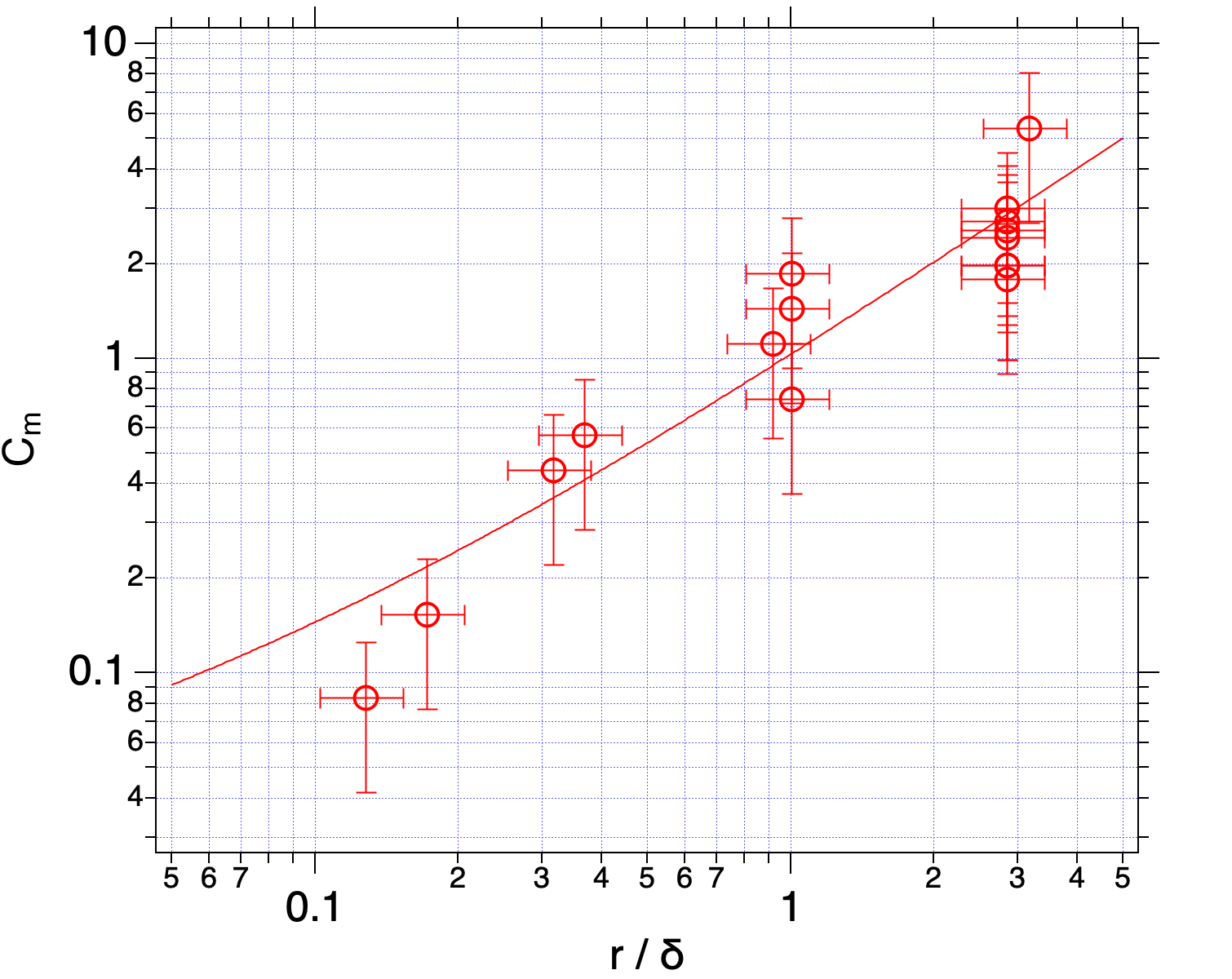}
  \includegraphics[scale=0.25]{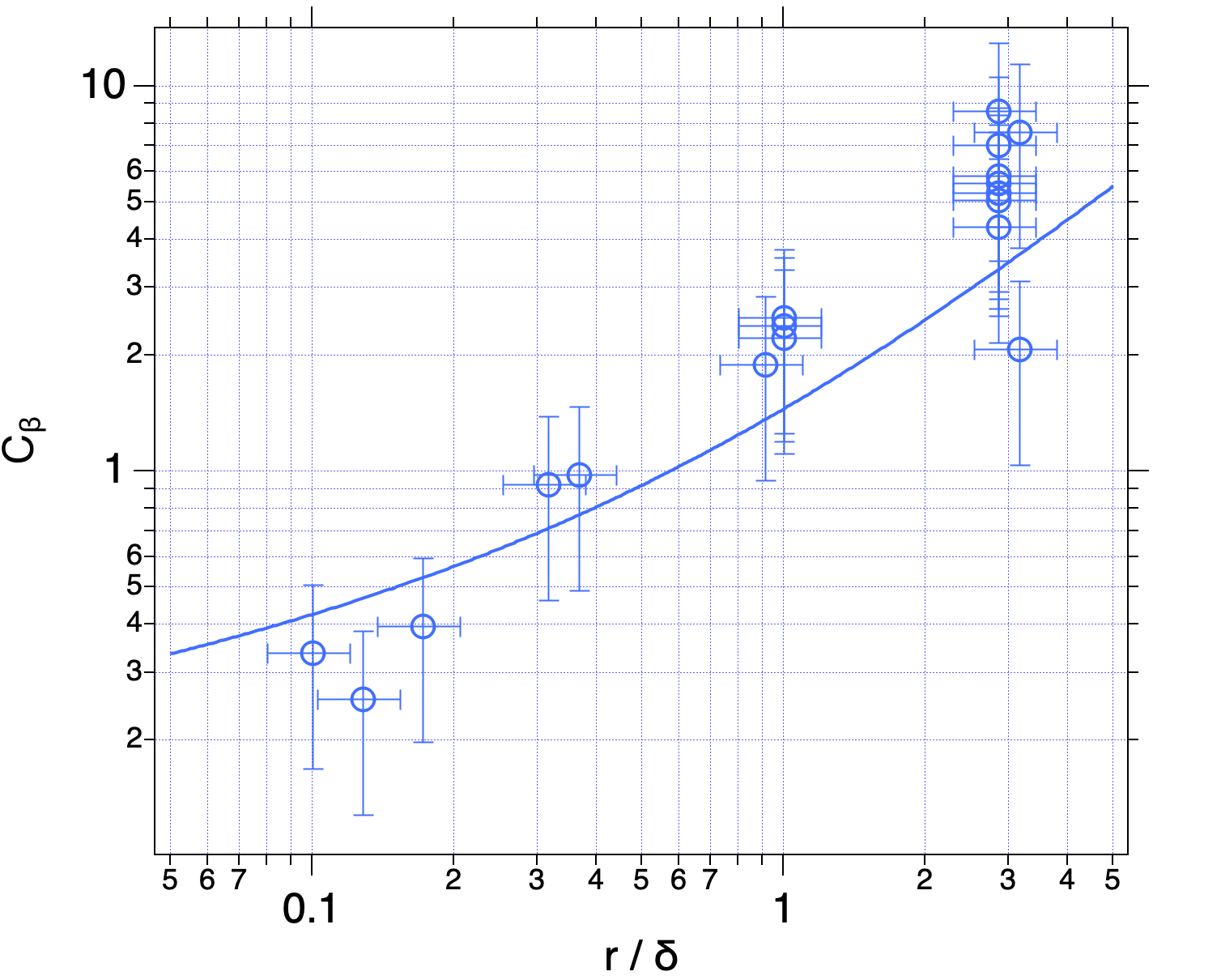}
  \caption{Dimensionless drag mass and friction coefficients determined as $C_m=\frac{\Delta m^* \omega}{2\pi \eta}$ and $C_{\beta}=\frac{\beta ^*}{2\pi \eta}$ as a function of $r/\delta$ for liquids of different viscosities and densities (listed in Table \ref{tbl:fluid_properties}). The solid lines correspond to the theoretical values of $C_m$ and  $C_{\beta}$ from the model. See text for details.}
  \label{sch:Cm_Cbeta}
\end{figure}

Figure \ref{sch:Cm_Cbeta} presents the measurements performed on the pure liquids of Table \ref{tbl:fluid_properties} in the 
form of dimensionless coefficients, that are compared to a master curve of $C_m$ and $C_{\beta}$ as a function of $r/\delta$ given by this model.

The viscosity of the fluids varies over 3 orders of magnitude, and the measurements reasonably reproduce the orders of magnitude of the model. Moreover, the measured values are in good agreement with those reported in the study by T. Ondarçuhu \textit{et al}\cite{DupredeBaubigny2016}, recorded using different setups at different amplitudes and frequencies of oscillation. This confirms  the reproducibility of the measurement protocol, allowing us to apply it to the study of the interface between two immiscible liquids. We test our setup on the widely studied PDMS / water interface, using a liquid PDMS with a kinematic viscosity of 10cSt ($10^{-5}m^2.s^{-1}$)

\subsection{Investigating the 10cst PDMS / water interface}

\begin{figure}
	\includegraphics[scale=0.25]{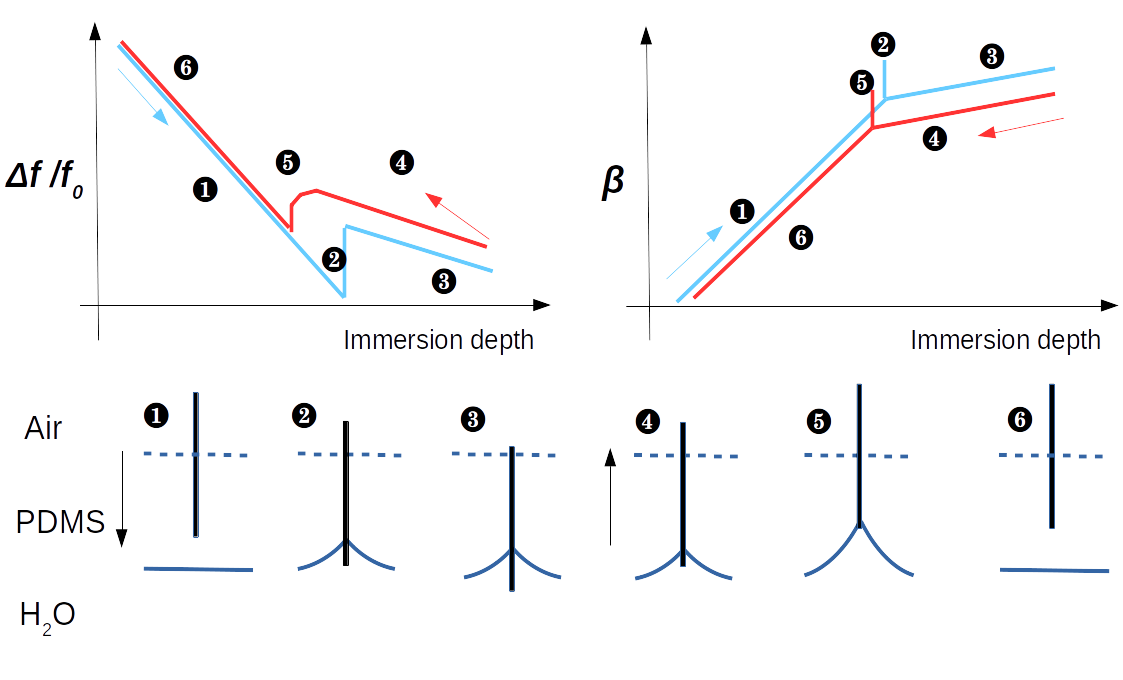}
	\caption{Schematic representation of signals of  frequency shift $\Delta f/f_0 $(left) and friction $\beta$ (in N.m.s$^{-1}$, right) with changes in the geometry of the fiber and interface when the force probe crosses the interface between two liquids of different viscosities. }
	\label{schm:schematic_app_ret}
\end{figure}

In terms of fluid properties,  an interface between two immiscible liquids such as siloxane and water, can be seen as a discontinuity in viscosity and density. The schematics of Figure \ref{schm:schematic_app_ret} depict the expected general shape of the approach / retract curves of frequency shift and dissipation signal. 

As the fiber approaches the PDMS / water interface (step 1), both the length of the fiber immersed in PDMS and the mass of the fluid dragged by the fiber increase.  Accordingly the frequency shift decreases linearly with a slope related to PDMS viscosity. 

At the interface crossing (step 2) the water forms a meniscus around the fiber resulting in an additional stiffness of the system \cite{Yazdanpanah2008,Jai2007}, and thus a positive frequency shift jump according to equation \ref{eq:freq_shift}. 

When the fiber moves in water (step 3), the frequency shift decreases linearly with a slope that depends on water viscosity. 

During retraction, when the fiber moves in water (step 4), the water dragged mass decreases. We thus expect to find the same absolute value for the slope of the frequency shift as in step 3. Frequency shift approach / retraction curves are slightly separated. This is to account for a possible hysteresis of the contact angle and/or a transient anchoring of the triple line on reversal of the movement.

At some point (step5) the contact line is pinned at the apex of the fiber, and the upward motion of the fiber causes the contact angle to decrease from its equilibrium value down to zero. When the contact angle has a zero value, the meniscus breaks.

Following this (step6 ), the frequency shift is expected to vary with the decreasing dragged mass of PDMS, up to the surface of the PDMS bath.

The same reasoning can \textit{a priori} be applied to the dissipation signals, with variations opposite to those of the frequency shift, and obviously different values for slopes in the linear portions of the curves. The only expected differences concern meniscus formation and breaking: the changes in conservative interactions (stiffness variations) are not expected to influence dissipation. However a specific dissipation excess can occur during the meniscus formation / breaking, as a result of discontinuities around the contact line \cite{Huh1971,Guo2013}, represented on the diagram as a sharp peak. 

Here, the measurements were performed on a millimetric drop of water placed at the bottom of the liquid cell filled with PDMS. As a rule, the interface was initially located at a depth of 150 to 200 microns below the PDMS surface. Measurements were made as soon as possible after the drop was deposited, typically within 15 minutes.
We used a QTFB sensor, with a fiber of radius $r=6.5 \mu$m ($f_0=32132.4$ Hz, $Q=7970$, $\beta _0=2.95.10^{-5}$N.m$^{-1}$.s). The vertical speed of the fiber was set to 0.5$\mu$m.s$^{-1}$. The bandwidth was set to 10Hz and 100Hz respectively for the PLL and the PID loops. The vertical oscillation amplitude of the fiber was set at $z=2$nm.

\begin{figure}
	\includegraphics[scale=0.3]{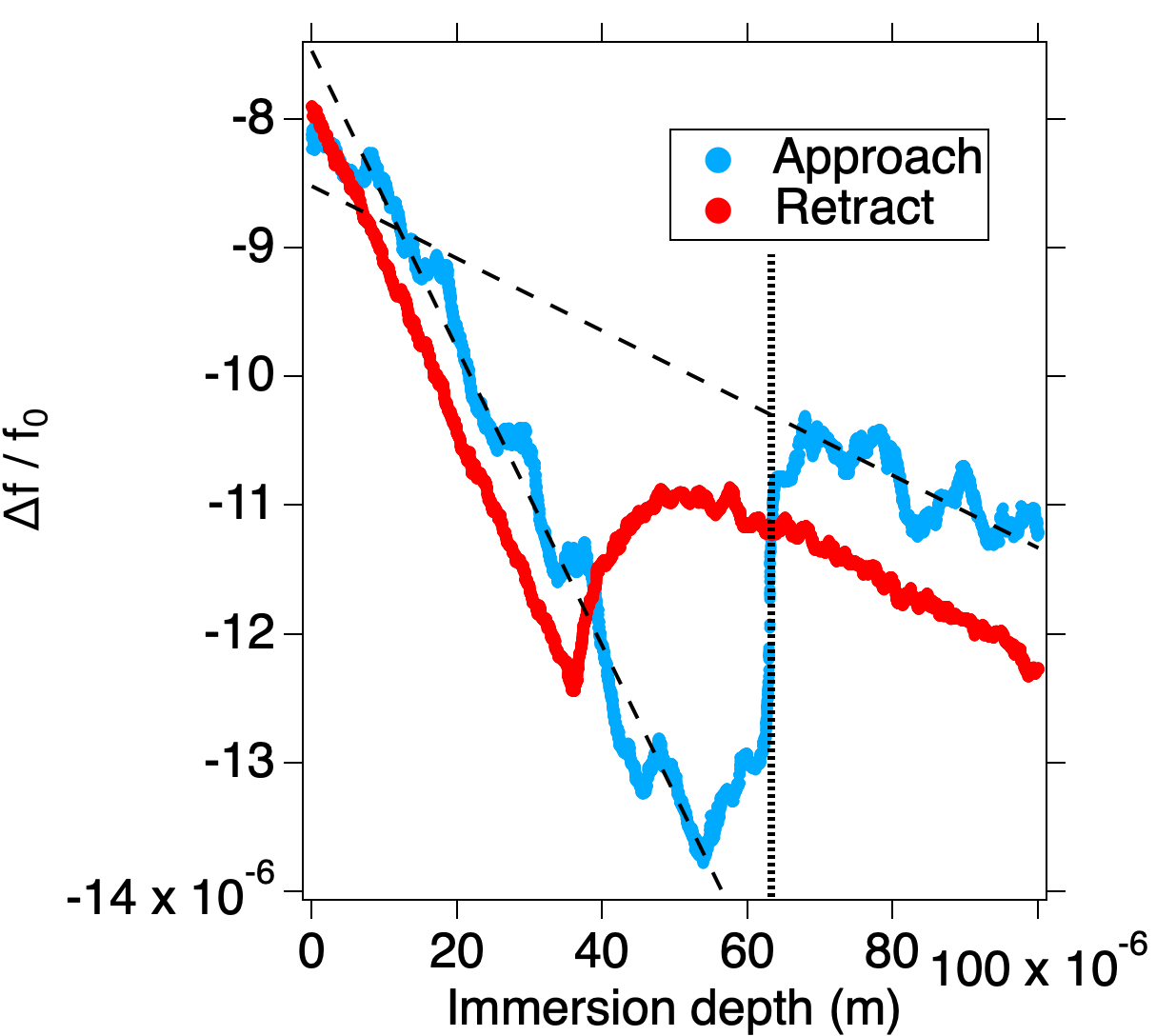}
	\includegraphics[scale=0.3]{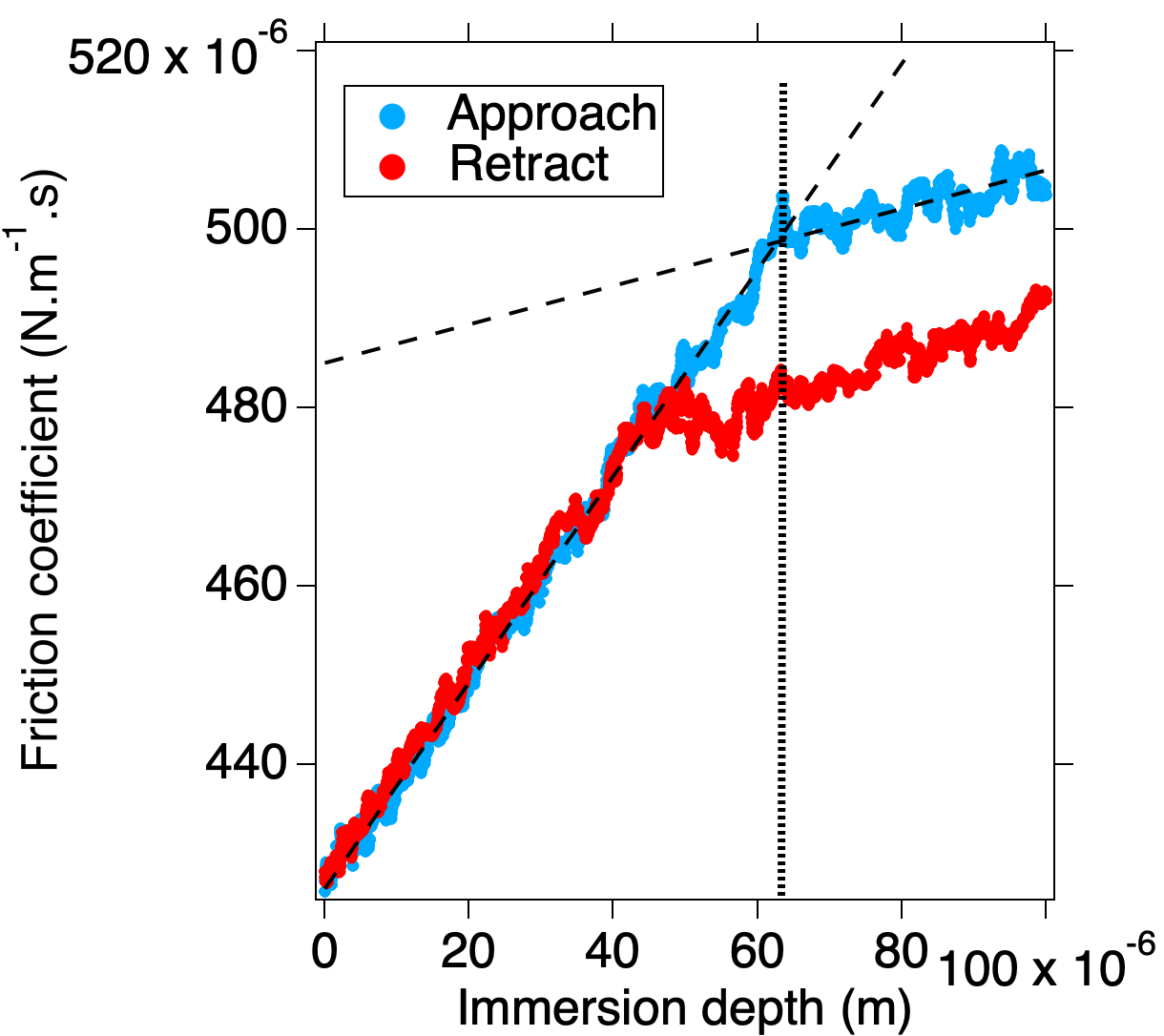}
	\caption{Simultaneous measurements of frequency shift and dissipation at the interface between 10cst PDMS oil and water using a QTFB with a fiber of radius $r=6.5 \mu$m, $f_0=32132.4$ Hz and $Q=7970$. Oscillation amplitude is $A_{RMS}=2$nm, and displacement occurs at a constant speed of 0.5$\mu $m.s$^{-1}$. The decrease in frequency shift with the immersion depth is due to the increase in mass of the fluid dragged by the fiber, while the positive contribution reflects the increased stiffness of the system arising from the water interfacial tension. The vertical dotted lines mark the depth at which the wetting of the fiber occurs during downward motion.}
	\label{schm:pdms10cst_h2o_app_ret}
\end{figure}

Figure \ref{schm:pdms10cst_h2o_app_ret} shows typical conservative and dissipative signals recorded during a forward and backward vertical motion of the oscillating fiber in the vicinity of the interface. The initial depth of the PDMS / water interface is 180$\mu$m for the data presented here. Data are presented in the form of a normalized frequency shift $\Delta f / f_0$ and a friction coefficient, $\beta$ in N.m$^{-1}$.s. The zero position on the graph does not correspond to the PDMS / air interface, but rather to the beginning of the signal recording. The curves exhibit the general behavior described by the diagrams in Figure \ref{schm:schematic_app_ret}.

During downward motion, frequency shift and dissipation signals vary linearly with depth, and the ratio of their slopes in PDMS and water corresponds to the ratio of the square root of the kinematic viscosities of PDMS (10cSt) and water (1cSt) as expected, with 21\% and 24\% error respectively for frequency shift and dissipation. Fiber wetting is observed at the immersion depth of 64$\mu$m, leading to a positive frequency shift. Frequency shifts and dissipation evolve linearly below this point, their slope depending on the viscosity of the water.

As the interface is approached, at an apparent separation of about 10 microns, a positive contribution to the frequency shift is measured, indicating an increase in the effective stiffness of the oscillator. As no change in the dissipation signal is observed, it can safely be assumed that this is an elastic effect. While the sudden increase is due to water wetting the fiber, we attribute the monotonic variation observed over several microns to the deformation of the oil/water interface. This deformation arises from the hydrodynamic force exerted by the probe on the interface \cite{Davis1989}. It is accompanied by a variation in the effective stiffness of the interface \cite{Horn1996}, and was described in earlier AFM studies \cite{Hartley1999,Aston2001}. The deformation caused by a drainage film between the probe and the interface is interpreted using the augmented Young-Laplace equation \cite{Manica2008,Bai2021}, with wetting occurring at the breakdown of the drainage film. In this regime the amplitude of the displacement of the fiber reflects the amplitude of the vertical deformation of the interface. Determining the distance between the fiber and the interface, i.e. the thickness of the drainage film, requires a modeling described in ref \cite{Manica2008}.  It is beyond the scope of this study, but can be estimated to a few hundred of nm.

During the upward motion of the fiber, while the fiber apex is in the water, the height of the fiber immersed in PDMS remains constant. Thus, the slope of frequency shift and dissipation measured is comparable to the slope measured during the downward motion, and again reflects the viscosity of the water. The breaking of the water meniscus and the dewetting of the fiber result in a negative frequency shift. It occurs at a depth of 36$\mu$m, above which, frequency shifts and dissipation continue to evolve linearly, with a slope that reflects PDMS viscosity. 

We estimate the height of the water meniscus by measuring the distance between the wetting and dewetting events, finding $h=27(\pm 1)\mu$m. Using the expression of James \cite{James1974} for the height of a meniscus on a cylinder, and taking a contact angle value and a capillary length for water in PDMS of respectively $45^\circ$ and 2.7mm  \cite{Svitova2002}, we obtain a meniscus height of 26$\mu$m in excellent agreement with our estimation.

Both frequency shift and dissipation approach / retraction curves are slightly separated. This is likely due to a transient anchoring (between 120 and 105 microns) of the triple line on reversal of the movement (and thus a reduction of the contact angle) on defects at the surface of the fiber probably resulting from the chemical thinning of the fiber. This is accompanied by an excess of dissipation and slightly decreased effective stiffness. In addition, the receding contact angle is expected to be lower than the advancing one \cite{Yazdanpanah2008}, which also contributes to the decrease in effective stiffness observed during the first part of upward motion.

It should be noted that the dissipation curves do not clearly exhibit excess of dissipation at the interface crossings. This suggests that any specific dissipation processes occuring in the meniscus are below the sensitivity threshold of our probe.

\subsubsection{Monitoring the displacement of the PDMS / water interface}

\begin{figure}
  \includegraphics[scale=0.3]{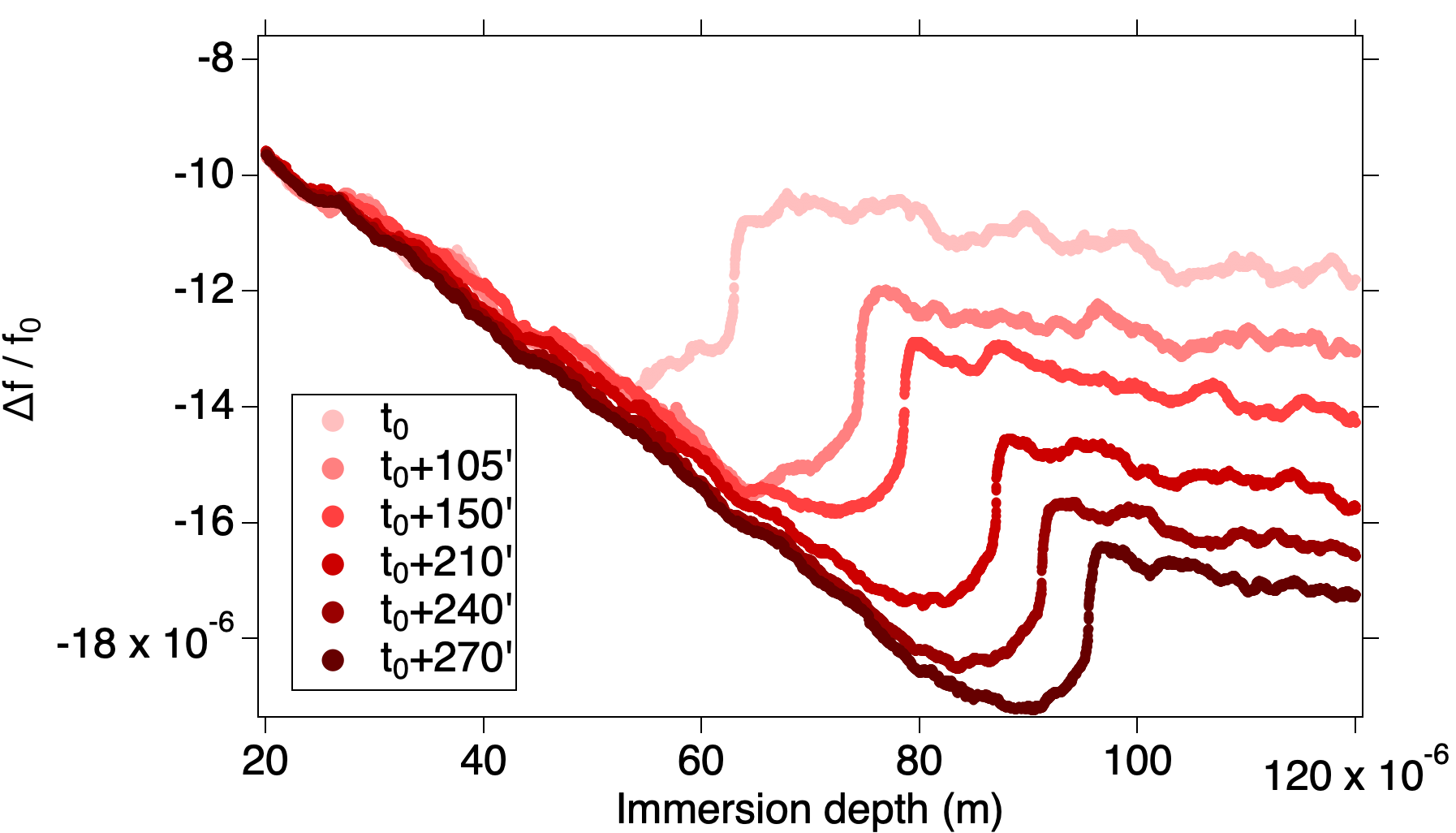}
  \includegraphics[scale=0.26]{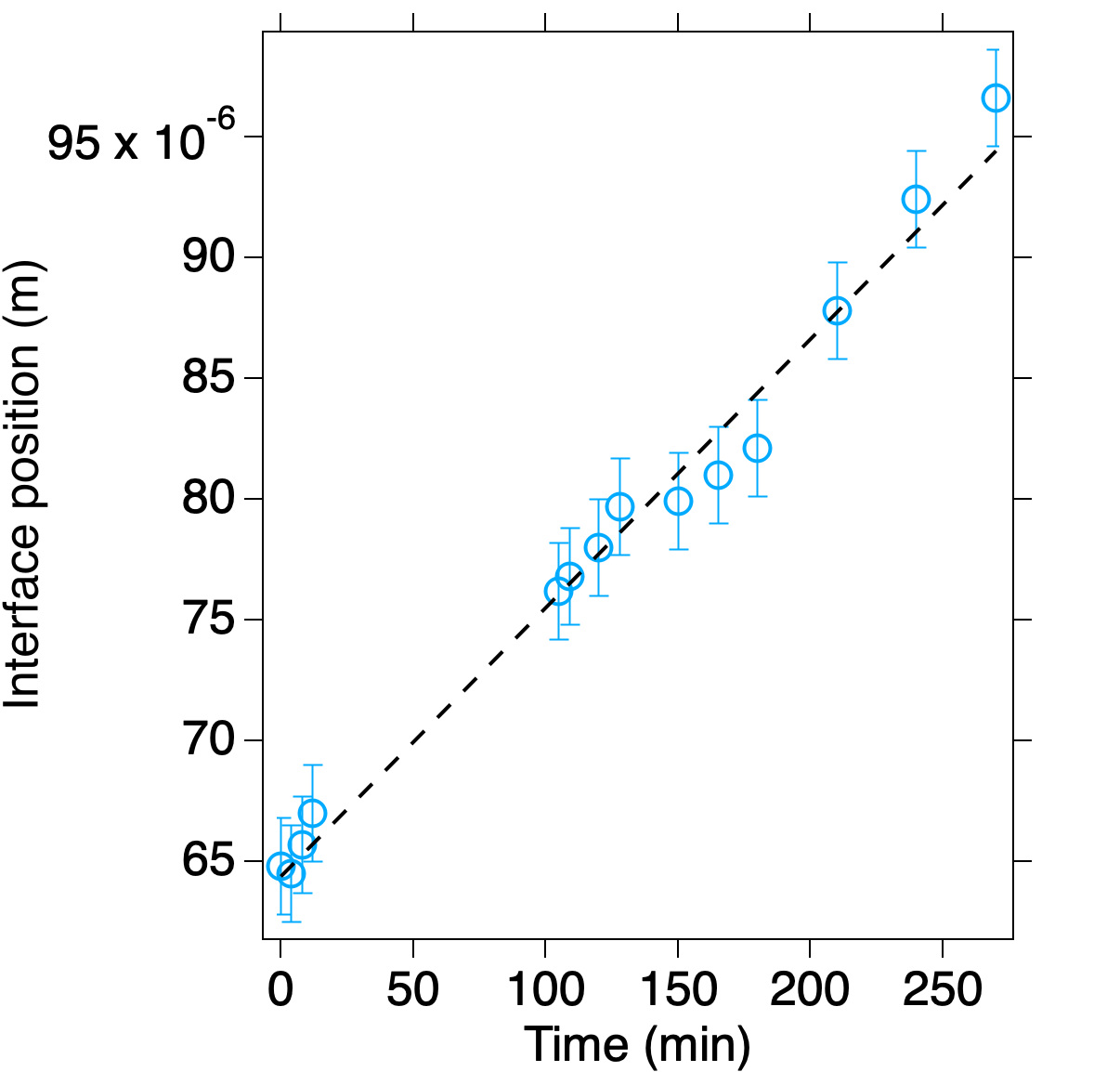}
  \caption{Monitoring of the interface depth with time over four hours following the protocol used to record data shown in Figure \ref{schm:pdms10cst_h2o_app_ret}. Left : As water evaporates into PDMS, the volume of the droplet decreases, causing the interface to shifts downward. To better visualize this displacement, the frequency shift curves have been surimposed along the frequency shift axis at the depth of 20 $\mu$m. Right : Plot of the interface position over time. The dashed line is the best linear fit giving an interface recession speed of $1.1 (\pm 0.1).10^{-7}$m.min$^{-1}$}
  \label{schm:deplacement_interface}
\end{figure}

We have so far considered that water is not soluble in PDMS for the sake of simplicity. However, while very low, water solubility in PDMS is not zero; it can be estimated to a few ppm by weight \cite{Vogel1964}. Actually, Grossier et al \cite{Rodriguez-Ruiz2013,Grossier2018} showed unambiguously that microdroplets of water immersed in liquid PDMS do evaporate.
Attempting to observe this phenomenon, we record numerous approach / retract curves like those in Figure \ref{schm:pdms10cst_h2o_app_ret}. The results of measurements over four hours on the same water droplet are presented in the left part of Figure  \ref{schm:deplacement_interface}, following the protocol detailed in the previous section, and using the same probe. For greater clarity, only a few representative frequency shift approach curves are plotted, and the curves have been surimposed along the frequency shift axis at the depth of 20 $\mu$m to better highlight the displacement of the interface. A vertical piezo slider operated in closed loop, is used to re-position of the fiber at the same initial depth with a resolution better than 100nm before each new recording. This depth corresponds to the zero shown for the depth axis in Figures \ref{schm:deplacement_interface} and \ref{schm:breaking_force}.

The first frequency shift curve labeled $t_0$ is that of  Figure \ref{schm:pdms10cst_h2o_app_ret} , and the others are comparable, with an offset in position and in frequency shift values. Not only do these observations prove the repeatability of the measurements, but the recession of the interface position is clearly visible, and shows a continuous shift to greater depths. 

In order to plot the evolution of this position with time, the depth at which the fiber reaches the water below the PDMS (as defined in the previous section) was recorded on each curve, with an uncertainty of 1 $\mu$m. On the right, Figure \ref{schm:deplacement_interface} plots of the evolution of the interface position with time during the experiment. It varies linearly, so that a linear fit yields a recession speed of $1.1(\pm 0.1).10^{-7}$m.min$^{-1}$.

We now check for any changes other than the position of the interface, as both position and $\Delta f$ values vary during the experiment. Shifting the approach $\Delta f$ curves on the immersion depth axis to superpose the point where the fiber enters the water, reveals  that the positive frequency shift attributable to water wetting the fiber decreases over time (see graph in  Supplementary information). This suggests a temporal evolution of the interface properties, likely a decrease in interfacial tension. Interfacial tension is very sensitive to the presence of contaminants at the interface, and such contamination over a duration of several hours in an open environment cannot be ruled out.

\subsubsection{Interfacial tension estimation}

We measured the interfacial tensions of the PDMS / water system using the pending drop method. The results are summarized in Table \ref{tbl:gamma}. The values of the interfacial tensions of water and PDMS are in agreemnt with the published values, and the value of the PDMS/water interfacial tension is median to those of the air/PDMS and air/water interfaces.

\begin{table}
  \caption{Summary of interfacial tension measurements using the pending drop method. Measurements were performed at T=21.5°C.}
  \label{tbl:gamma}
  \begin{tabular}{cc}
    Interface & Interfacial tension (mN.m$^{-1}$) \\
    \hline
    Air / water & $72.1\pm 0.6$\\
    Air / PDMS & $20.4 \pm 0.2$\\
    PDMS / water & $37.1\pm 1.2$\\
    \hline
  \end{tabular}
\end{table}

\begin{figure}
  \includegraphics[scale=0.27]{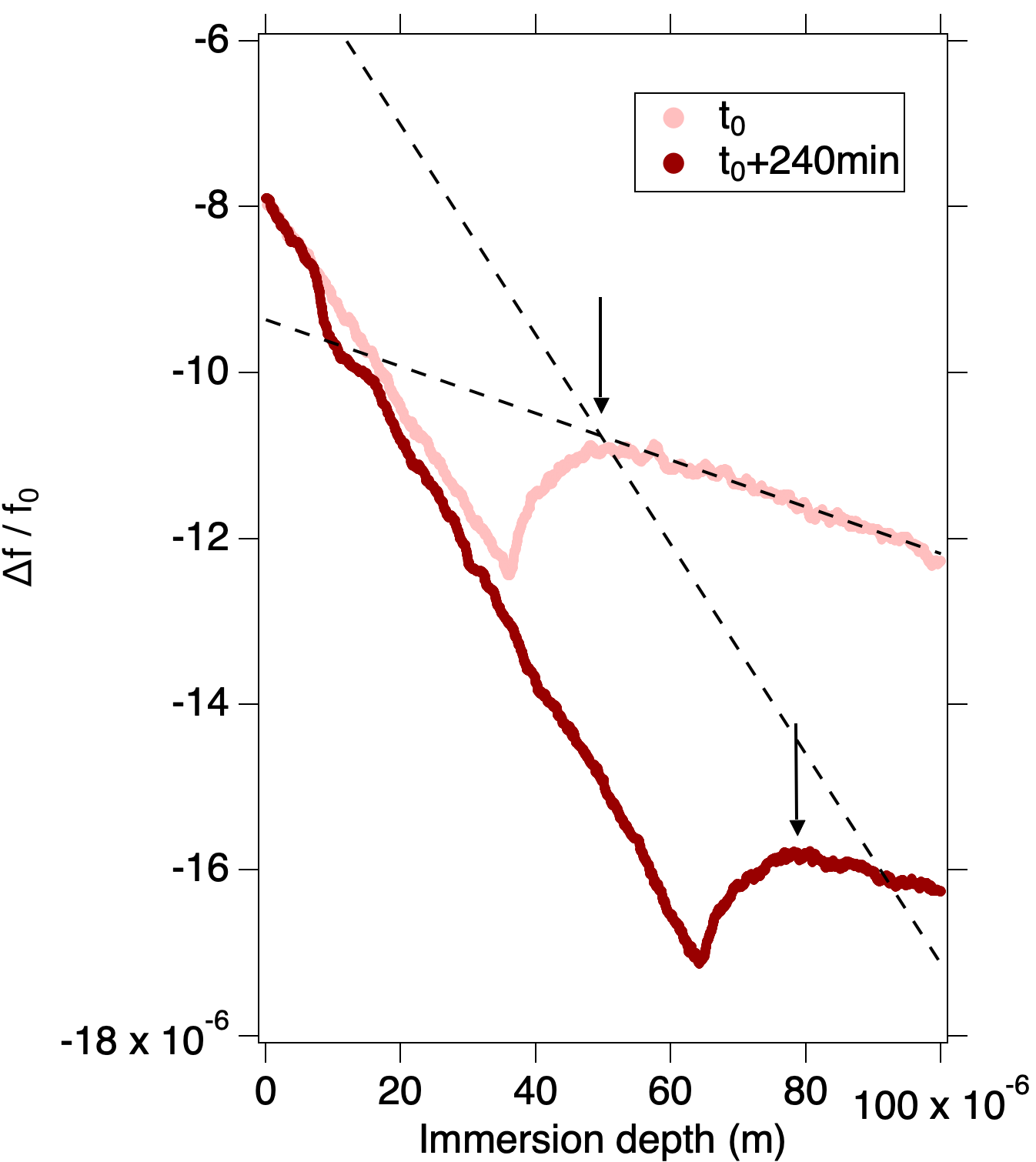}
  \includegraphics[scale=0.27]{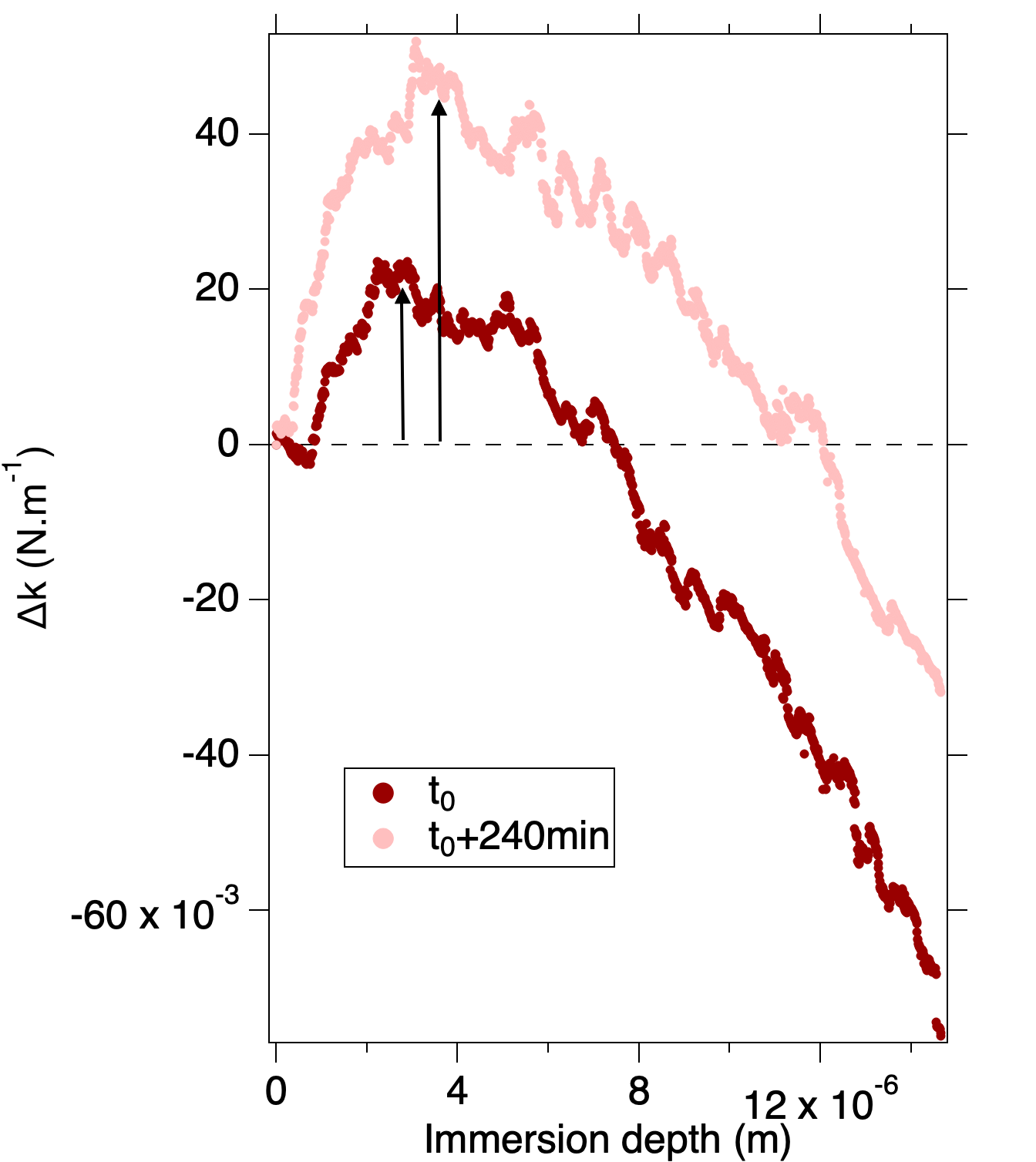}
  \caption{Plot of the frequency shift (left) and effective stiffness (right) at the PDMS water interface at $t_0$ and $t_0+240$min during upward motion of the oscillating fiber. The point where the water meniscus is pinned at the apex of the fiber is marked by arrows. The effective stiffness is proportional to interfacial tension. To calculate it, the dragged mass contribution of PDMS (linear fit plotted as dashed lines at $t_0$) to the frequency shift is first subtracted from the data, above the depth marked by arrows. We associate its maximum value with the meniscus stiffness. From this value, with a liquid tank radius $L=8.7$mm, and a fiber radius $r=6.5 \mu$m, we found an interfacial tension $\gamma=60(\pm 7)$mN.m$^{-1} $ at $t_0$, and $\gamma=29( \pm 3)$mN.m$^{-1}$ at $t_0+240$min. See text for details.}
  \label{schm:breaking_force}
\end{figure}

The interfacial tension of the PDMS / water interface can also be estimated from the frequency shift curves during the upward motion of the fiber. During retraction, when the meniscus gets pinned at the apex of the fiber, the receding contact angle will decrease down to zero upon breaking of the meniscus, thereby altering the effective stiffness of the oscillator.  This evolution is visible on the retraction frequency shift curves shown in the left graph of Figure \ref{schm:breaking_force} (at $t_0$ and $t_0+240$min). It appears  as a sharp decrease in frequency shift from the point marked by an arrow, as the fiber moves upward upon the water meniscus breaking. 

The frequency shift then varies linearly again, with a slope corresponding to a dragged mass of PDMS.
Prior to the breakup, the effective stiffness variation $\Delta k$ of the oscillator can be extracted using equation \ref{eq:freq_shift}, by removing the contribution of the dragged mass of the PDMS from the point marked by an arrow. The evolution of $\Delta k$ is plotted on the right graph of Figure \ref{schm:breaking_force}, the zero depth and stiffness corresponding to the breakup of the water meniscus. At $t_0$ and $t_0+240$min, we clearly observe that the stiffness reaches a maximum value, which we associate with the meniscus stiffness. 

Using the work of T. Ondarçuhu group on the effective spring constant of liquid interfaces \cite{DupredeBaubigny2015}, we can extract from it an estimate of the interfacial tension $\gamma _{int} \cong \Delta k \ln (L/r) / 2 \pi$, $L$ being the radius of the tank containing the liquid, and $r$ the radius of the fiber.

This analysis began with frequency shifts recorded at the PDMS / air interface (see Supplementary information), to obtain an estimated PDMS surface tension of  $\gamma=18(\pm2)$mN.m $^{-1}$ in agreement  with the measured values of the surface tension of PDMS.

At the PDMS / water interface, we estimate the interfacial tension at $\gamma_{int}=60(\pm 7)$N.m$^{-1} $ at time $t_0$. Interestingly, it decreases to $\gamma_{int}=29( \pm 3)$N.m$^{-1}$ at time $t_0+240$min. This reduction, larger than the experimental errors \bibnote{Error here is the experimental error estimated at 11\%, all surface tension measurements being carried out with the same probe. Total error including systematic errors is estimated at 30\%, which should be used to compare our interfacial tension estimations with literature values},  is consistent with a contamination of the interface  accompanied by decreased interfacial tension  \cite{Azizian2003}. It should be noted that a measurable decrease in interfacial tension at the oil / water interface \cite{Hartley1999} was reported in an earlier AFM study, in the presence of trace surfactants at the interface. The estimated interfacial tension is in agreement with the results obtained by the pending drop method and the previously published values \cite{Svitova2002,Ismail2009}.

The amplitude of the interface deformations apparent on the approach frequency shift curves (shown in Supplementary information), and discussed above, depends on the interfacial tension. A decrease in interfacial tension must be accompanied by a decrease in the amplitude of the deformation before wetting. We indeed observe a decrease in the distance traveled by the fiber  from the point where an elasto-hydrodynamic contribution is measured to the point where wetting occurs. The value of $9.5 \pm 0.5 \mu$m measured at time $t_0$ decreases to $6.5 \pm 0.5\mu$m at $t_0+270$ min, consistent with the evolution of interfacial tension.

Taken together, these preliminary results clearly show that we are able to probe an interface between two fluids with our hanging fiber AFM. In addition to being quantitatively sensitive to viscosity change, this setup enables us to observe the wetting of the fiber and thus to estimate the interfacial tension. It also enables us to locate the interface and monitor its displacement while water 'evaporates' into the oil. Thanks to the sensitivity of the probe, the interface's elasto-hydrodynamic response is measured as the fiber approaches it, before wetting occurs. This non-contact interaction with the interface, combined with the FM-AFM operating mode, opens the way to locally measuring the elastic and dissipative properties of the interface at a constant distance over time.

Although the sensitivity of the force measurements is significantly improved compared to conventional AFM probes, the technique has some limitations. The deeper the interface, the more the sensitivity of the measurements is reduced due to the damping of the oscillator, leading to a strong decrease of the quality factor of the probe. The same factor also limits the viscosity of the fluids that can be studied with these probes. Moreover, during the measurements, the localization of the interface is linked to the existence of a density and/or viscosity gradient, as well as to the wetting of the fiber when crossing the interface. It appears difficult to apply this technique to interfaces that do not meet these specifications.
Regarding the lateral resolution of this type of measurements, the limitation comes from the skin thickness of the moving fluid layer. In the case of the results presented here, it is in the range 3-10 microns, but is completely dependent on the kinematic viscosity of the fluids.

It should be noted that the probe as presented here, is not the most suitable to study the deformation of an interface, because the shape of the fiber apex is not well defined. We plan to use a spherical end fiber for future studies, for example by attaching a coloidal silica sphere at the end of the fiber, in order to model the contact geometry correctly.

\section{Conclusion}

This paper presents a method of measuring the interaction forces in a simple fluid, or at an interface between two liquids using an original FM-AFM in a hanging fiber geometry.

After detailing the manufacture and calibration of this very high sensitivity force probe based on a cheap watch quartz tuning fork,  we demonstrate its ability to make quantitative and reproducible measurements of simple fluid properties. 

Furthermore, we provide evidence that this AFM is suitable for measurements at the interface between two liquids. In particular, the repeatability of the measurements allows a moving interface to be monitored over time. The evaporation of a millimetric drop of water immersed in PDMS is observed, yielding a reasonable estimate of the interfacial tension and highlighting its decrease over time due to contamination. 

These capabilities, and  preliminary results on the interface between two immiscible liquids, pave the way for interface manipulation, such as triggering crystallization in liquid micro-reactors \cite{Grossier2011}. Moreover, we also show that non-contact interactions can be envisaged, in the spirit of the work of E. Charlaix's on using elasto-hydrodynamic interaction for the non-contact measurement of elastic properties of thin films  \cite{Steinberger2008,Leroy2011}. This technique could be fruitfully applied to the study of complex fluid-fluid interfaces \cite{Fuller2012,Jaensson2018}. But also in the field of biology, to the mechanical properties of membranes, as well as the the phenomena of locomotion exploiting the deformation of interfaces \cite{Trouilloud2008}.

\section{Abreviations}

\begin{table}
  \caption{Abbreviations used in the text}
  \label{tbl:abbrev}
  \begin{tabular}{ll}
  AFM & Atomic Force Microscope\\
  FM-AFM & Frequency-Modulation Atomic Force Microscope\\
  FWHM & Full Width at Half Maximum\\
  L  & Length  \\
  PDMS & Poly-DiMethylSiloxane\\
  PID & Proportional-Integral-Differential\\
  PLL & Phase-Locked-Loop\\
  Q & Quality factor\\
  QTF & Quartz Tuning Fork\\
  T & Thickness \\
  W & Width \\
  \end{tabular}
\end{table}

\begin{acknowledgement}

The authors thank Dr T. Ondarçuhu for helpful discussion concerning the quantitative rheological measurements, M. Lagaize for his crucial technical support (design and production of mechanical parts, 3D printing), and D. Chaudanson for the SEM imaging of the QTF probe.

\end{acknowledgement}

\begin{suppinfo}

\begin{itemize}
 \item photographs and description of the AFM setup
 \item protocol for the mechanical and electrical calibration of bare QTFs
 \item evolution of the frequency shift curves at the approach of the PDMS / water interface over time
 \item frequency shift and dissipation measurements at air / PDMS interface
\end{itemize}

\end{suppinfo}

\bibliography{ms}

\begin{mcitethebibliography}{59}
\bibitem[Not()]{Note-2}
Error here is the experimental error estimated at 11\%, all surface tension
  measurements being carried out with the same probe. Total error including
  systematic errors is estimated at 30\%, which should be used to compare our
  interfacial tension estimations with literature values\relax
\mciteBstWouldAddEndPuncttrue
\mciteSetBstMidEndSepPunct{\mcitedefaultmidpunct}
{\mcitedefaultendpunct}{\mcitedefaultseppunct}\relax
\EndOfBibitem
\end{mcitethebibliography}

\end{document}